\documentclass[twocolumn,prl,showpacs]{revtex4}
\usepackage{graphicx}
\usepackage{dcolumn}
\usepackage{bm}
\begin{document}

\title{Accelerating expansion of the universe may be caused by inhomogeneities}
\author{Gyula Bene$^{1}$\footnote{Electronic address: bene@arpad.elte.hu}, Viktor Czinner$^{2}$\footnote{Electronic address: czinner@rmki.kfki.hu} and M\'aty\'as Vas\'uth$^{2}$\footnote{Electronic address: vasuth@sunserv.kfki.hu}}
\affiliation{
$^1${\it Institute for Theoretical Physics, E\"otv\"os University,
     P\'azm\'any P\'eter s\'et\'any 1/A, H-1117 Budapest, Hungary,}\\
$^2${\it KFKI Research Institute for Particle and Nuclear Physics, Theory Department
H-1525 Budapest, P.O.Box 49, Hungary}
}
\date{\today}

\begin{abstract}
We point out that, due to the nonlinearity of the Einstein equations,
a homogeneous approximation in cosmology leads to the appearance
of an additional term in the Friedmann equation. This new term is associated with
the spatial inhomogeneities of the metric and can be expressed in terms of density
fluctuations. Although it is not constant, it decays much slower (as $t^{-\frac{2}{3}}$)
than the other terms (like density) which decrease as  $t^{-2}$. The presence of the new term
leads to a correction in the scale factor that is proportional to $t^{2}$ and may give account
of the recently observed accelerating expansion of the universe without introducing a cosmological constant.
\end{abstract}


\maketitle


Measurements of the light curves of several hundred type Ia supernovae \cite{Schmidt:1998,Riess:1998cb, Perlmutter:1998np, Tonry:2003zg} 
and other independent observations \cite{Bennett:2003bz,Netterfield:2001yq,Halverson:2001yy,Szalay} convincingly demonstrate that the expansion 
of the universe is accelerating.  

This unexpected result stimulated a number of theoretical investigations. Most explanations suggested so far
seem to belong to one of three categories: assuming a nonzero cosmological constant \cite{Carroll:2000fy,Peebles:2002gy},
assuming a new scalar field (``quintessence'') \cite{Wetterich:fm,Ratra:1987rm,Caldwell:1997ii,Armendariz-Picon:1999rj,Armendariz-Picon:2000dh,Armendariz-Picon:2000ah,Mersini:2001su,Caldwell:1999ew,Carroll:2003st,Sahni:1999gb,Parker:1999td}, 
or assuming new gravitational physics \cite{Deffayet:2001pu,Freese:2002sq,Ahmed:2002mj,Arkani-Hamed:2002fu,Dvali:2003rk,Turner}.


In the present paper we follow a fourth approach, namely, we attribute the accelerating expansion to a
consequence of the inhomogeneities present in ordinary matter, according to the dynamics described by the 
usual Einstein equations. A similar approach has been suggested in \cite{negfric1}.
We assume a matter dominated and (on the average) flat universe 
(zero pressure, $\Omega=1$, $k=0$), without cosmological constant or any other exotic constituent. 

It is natural and
usual to apply a homogeneous approximation for the metric, since universe is indeed 
homogeneous on the large scale. 
 Explicitly, this means that
instead of the actual space dependent metric $g_{ik}(t,\bm r)$ one uses its spatial average,
i.e.
\begin{eqnarray}
\overline{g_{ik}}(t)=\lim_{V\rightarrow \infty} \frac{1}{V}\int_V d^3 \bm r\; g_{ik}(t,\bm r)\; .\label{f1}
\end{eqnarray}  

The actual, not precisely homogeneous metric $g_{ik}(t,\bm r)$ satisfies Einstein's equations
\begin{eqnarray}
R_{ik}(\{g_{ik}\})-\frac{1}{2}g_{ik} R(\{g_{ik}\})=\frac{8\pi G}{c^4} T_{ik}\; .\label{f2}
\end{eqnarray} 

Since these equations are nonlinear and since there are strong inhomogeneities on smaller scales,
the spatially averaged homogeneous metric (\ref{f1}) together with the spatially averaged matter density 
will not satisfy Eq.(\ref{f2}). In other words,
inhomogeneities (which, unlike a cosmological constant or quintessence, are unquestionably present)
induce a correction term in the Friedmann equation. Such a procedure 
has been introduced for the case of gravitational radiation in \cite{pertx}, applied for an ideal fluid
in  \cite{pert2} and for the case of a scalar field in \cite{pert1}.  We demonstrate this below in the framework of
second order perturbation theory. First order corrections are well known \cite{Lifsic,Lifsic-Khalat,Sachs-Wolfe}.
We write down the spatial average of the second order perturbation equations, which (as they are linear
in the second order correction of the metric and the density) enable us to calculate the spatial average
of the metric and the density.


The metric is written in the form
\begin{eqnarray}
g_{jk}=g_{jk}^{(0)}+g_{jk}^{(1)}+g_{jk}^{(2)}\; ,\label{e1}
\end{eqnarray}
where the upper, bracketed indices refer to the order of the perturbation.
Henceforth we use comoving coordinates. Assuming isotropy, we have
\begin{eqnarray}
\overline{g_{00}}(t)=c^2\; ,\label{e3}\\
\overline{g_{0\alpha}}(t)=0\; ,\label{e4}\\
\overline{g_{\alpha\beta}}(t)=-R^2(t)\delta_{\alpha\beta}\; ,\label{e5}
\end{eqnarray}
where the bar over quantities means spatial average (cf. Eq.(\ref{f1})). Especially, we have
\begin{equation}
\overline{g_{\alpha\beta}^{(i)}} = -\delta_{\alpha\beta}\left(R^2\right)^{(i)}(t) \ , \qquad i=0,1,2\; .\label{e6}
\end{equation}

The use of comoving coordinates implies that
\begin{eqnarray}
T^{00}=\rho=\rho^{(0)}+\rho^{(1)}+\rho^{(2)}\; ,\\
T^{0\alpha}=0\; ,\\
T^{\alpha\beta}=0\; .
\end{eqnarray}
Zeroth order quantities are those in a flat Friedmann model with matter domination, i.e.

\begin{eqnarray}
R^{(0)}&=&\frac{2c}{H_0}\left(\frac{3}{2}H_0t\right)^{\frac{2}{3}}\;,\\
\rho^{(0)}&=&\frac{1}{6\pi G t^2}\;.
\end{eqnarray}

First order corrections are given by \cite{Sachs-Wolfe}
\begin{eqnarray} 
 g^{(1)}_{\alpha \beta} &=& \frac{4c^2\eta^4}{H_0^2}\left[{ 1 \over \eta}\; 
{ \partial \over \partial  \eta}\; \left({ 1 \over \eta}\; 
 D_{\alpha \beta} \right)\right.\nonumber\\\left.\right. &&\left.\;  - 2 \left({ 8 \over  \eta^3} 
-{ \Delta \over \eta} \right)( C_{\alpha,\beta } +C_{\beta,\alpha} ) \right.  \nonumber\\ 
 \left.\right. &&\left.\;+{ A_{ ,\alpha\beta}\over  \eta^3}+\eta_{\alpha \beta} B - 
 { \eta^2\over  10} B_{ , \alpha\beta}\right] \; ,    \label{p1}        \\ 
  g^{(1)}_{\alpha 0} &=& -\frac{4c^2}{H_0} \Delta C_{\alpha}\; ,   \label{p2} \\ 
 \rho^{(1)}\; &=&\; { H^2_0\over  32\pi G}\; 
\Delta \; \left(\; { 6A\over \eta^9} -{3\over 5} 
{ B \over  \eta^4}\;\right)  \label{p3} 
\end{eqnarray} 
with
\begin{eqnarray}
\eta=\left(\frac{3}{2}H_0t\right)^{\frac{1}{3}}\; .
\end{eqnarray}
Here $A$, $B$ and $C_{\alpha}$ are functions depending only on the spatial coordinates. $A$ and $B$ are arbitrary 
(determined by the initial conditions) while $C_{\alpha}$ has zero divergence. Finally, $D_{\alpha \beta}$ satisfies a wave equation.
Note that the assumption of isotropy implies $C_{\alpha}=0$ and $D_{\alpha \beta}=0$.
Henceforth we set $A=0$, too.

Second order corrections of the (0,0) component of the Einstein equations have the general structure

\begin{eqnarray}
&& (\rm{terms\quad linear\quad  in\quad} g^{(2)}_{ik})\nonumber\\
&+&(\rm{terms\quad  quadratic\quad  in\quad}  g^{(1)}_{ik})=8\pi G\;\rho^{(2)}\; .
\end{eqnarray}

Calculating the spatial average of this equation we have

\begin{eqnarray}
&& 6\frac{\dot {R}^{(0)}}{{R}^{(0)}}\left(\frac{\dot {R}^{(2)}}{{R}^{(0)}}-\frac{\dot {R}^{(0)}R^{(2)}}{\left({R}^{(0)}\right)^2}\right)+\frac{1}{400}H_0^2\;\overline{\left(\triangle B\right)^2}\;\frac{1}{\eta^2}
\nonumber\\
&&-\frac{17}{80}H_0^2\;\overline{\left(\bm \nabla B\right)^2}\;\frac{1}{\eta^4}
 =8\pi G\;\rho^{(2)}\; .
\end{eqnarray}

Similarly, for the spatial average of the second order corrections of the  0 component 
of the divergence equation we obtain

\begin{eqnarray}
&&\frac{1}{\left(R^{(0)}\right)^3}\left(\rho^{(2)}\left(R^{(0)}\right)^3\right)_{,0}
-\frac{1}{2}\overline{\rho^{(1)}\left(\frac{g^{(1)}_{\alpha\alpha}}{\left(R^{(0)}\right)^2}\right)_{,0}}
\nonumber\\
&+&\frac{1}{2}\rho^{(0)}\left\{3\left(2\frac{R^{(2)}}{R^{(0)}}+
\frac{1}{4}\left(\overline B\right)^2\right)_{,0}
-\frac{1}{\left(R^{(0)}\right)^4}\overline{g^{(1)}_{\alpha\beta}g^{(1)}_{\alpha\beta,0}}
\right.\nonumber\\\left.
\right.&-&\left.\frac{2\dot R^{(0)}}{\left(R^{(0)}\right)^3}\left[\frac{1}{c^2}
\overline{\left(g^{(1)}_{0\alpha}\right)^2}-
\frac{1}{\left(R^{(0)}\right)^2}\overline{\left(g^{(1)}_{\alpha\beta}\right)^2}\right]\right\}=0\; .
\end{eqnarray}

Its solution sounds
\begin{eqnarray}
8\pi G \rho^{(2)} =- 6\pi G\rho^{(0)}\left(4\frac{R^{(2)}}{R^{(0)}}+A_1
\right)\nonumber\\-\frac{3}{20}H_0^2\;\overline{\left(\bm \nabla B\right)^2}\;\frac{1}{\eta^4}
+\frac{9}{800}H_0^2\;\overline{\left(\triangle B\right)^2}\;\frac{1}{\eta^2}
\end{eqnarray}
where $A_1$ is an integration constant.
Putting this into the Einstein equation we get an ordinary differential equation for $R^{(2)}$
\begin{eqnarray}
\dot R^{(2)} +\frac{1}{3t}R^{(2)}&=&-\frac{3}{4}\frac{c\;A_1}{\left(\frac{3}{2}H_0t\right)^{\frac{1}{3}}}
+\frac{1}{48}\overline{(\bm \nabla B)^2}c\left(\frac{3}{2}H_0t\right)^{\frac{1}{3}}\nonumber\\
&+&\frac{7}{1600}\overline{(\triangle B)^2}H_0ct
\end{eqnarray}
with the solution
\begin{eqnarray}
R^{(2)}&=&A_2\;t^{-\frac{1}{3}}-\frac{1}{4}A_1 R^{(0)}
+\frac{1}{120}\frac{c}{H_0}\overline{(\bm \nabla B)^2}\left(\frac{3}{2}H_0t\right)^{\frac{4}{3}}\nonumber\\
&+&\frac{3}{1600}\overline{\left(\triangle B\right)^2}\;H_0\;c\; t^2\; .\label{xxx}
\end{eqnarray}

\begin{figure}[h]
\begin{flushleft}
\includegraphics{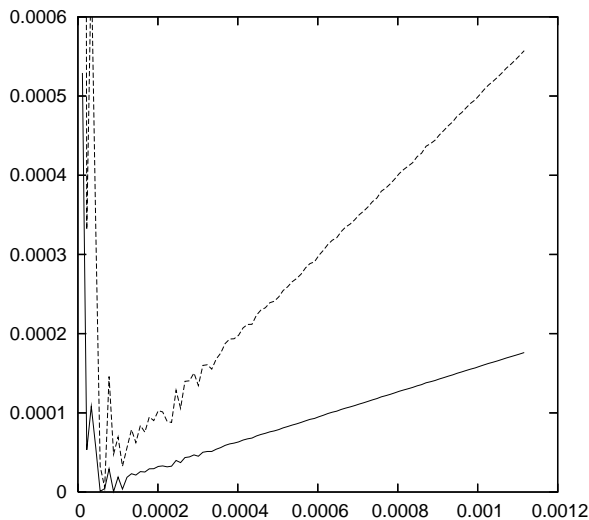}
\end{flushleft}
\caption{Relative error of the coefficients of Eq.(\ref{xxx}) versus $\sqrt{\overline{(\delta \rho /\rho)^2}}$. 
Solid line: relative error of the coefficient of the $t^2$ term,
dashed line: relative error of the coefficient of the $t^{4/3}$ term.}
\label{fig1}
\end{figure}

The integration constants $A_1$, $A_2$ are to be determined from the initial conditions for the second order correction of the 
metric and its time derivative. 
It is interesting to compare Eq.(\ref{xxx}) with the perturbation series of a homogeneous
open universe, by considering the curvature term as a perturbation in a flat universe:
\begin{eqnarray}
R_{o}=R^{(0)}+\frac{1}{10}\frac{c}{H_0}\left(\frac{3}{2}H_0t\right)^{\frac{4}{3}}
-\frac{27}{5600}\;H_0\;c\; t^2+...\label{xopen}
\end{eqnarray}
In order to check Eq.(\ref{xxx}), we determined numerically 
the second time derivative of the spatial metric from the Einstein equations by using an effective 
pseudospectral scheme. The relative error of the coefficients (mainly due to third order corrections)
is displayed in Fig.\ref{fig1}.

For sub-horizon-sized 
perturbations in the late universe the last term of Eq.(\ref{xxx}) is dominant.
The complete scale factor up to second order then reads 

\begin{eqnarray}
R(t)&=&R^{(0)}+\frac{3}{1600}\overline{\left(\triangle B\right)^2}\;H_0\;c\; t^2\nonumber\\
&=&R^{(0)}\left[1+\frac{1}{6}\overline{\left(\frac{\delta\rho}{\rho}\right)^2} \right]\; ,
\label{scale}
\end{eqnarray}
where $\delta\rho\equiv \rho^{(1)}$.

The present value of the fractional density fluctuation can be estimated as (provided
that linear perturbation theory is applicable, for $\frac{\delta\rho}{\rho}>1$ fluctuations grow faster)
\begin{equation}
\overline{\left(\frac{\delta\rho}{\rho}\right)^2_0} =\overline{ \left(\frac{\delta\rho}{\rho}\right)^2_{dec}} (1+z_{dec})^2 \approx 1...100\;,
\end{equation}
because
\begin{equation}
1+z_{dec}= \frac{R(t_0)}{R(t_{dec})} \approx 1100
\end{equation}
and
\begin{equation}
\left(\frac{\delta\rho}{\rho}\right)_{dec}\approx 10^{-2}...10^{-3}
\end{equation}
(see e.g. \cite{Kolb}).
Here $0$ and $dec$ stand for present value and value at decoupling, respectively. Thus, the final results are expressed in terms of the (large) relative density fluctuations. Eq.(\ref{scale}) demonstrates that taking into account of density 
fluctuations is essential in the late universe. A perturbative treatment may not even be sufficient. It is also seen that
the additional term due to inhomogeneities is a quadratic function of the time, thus for a long time after the Big Bang,
accelerating expansion occurs. This can be quantified in terms of the deceleration parameter which reads
 
\begin{equation}
q= -\frac{\ddot{R}R}{\dot{R}^2}= \frac{1}{2}\left[1-\frac{7}{3}\overline{\left(\frac{\delta\rho}{\rho}\right)^2} \right]\; .
\end{equation}

Thus, according to second order perturbation theory the deceleration parameter becomes negative (acceleration)
if $\delta\rho /\rho>\sqrt{3/7}\approx 0.65$. Assuming that $\overline{\left({\delta\rho}/{\rho}\right)^2_0} \approx 1 $,
this happens in our approximation at $z=\sqrt{7/3}-1\approx 0.53$.  

It is instructive to calculate the effective pressure as well. In a homogeneous universe with the scale factor
(\ref{scale}) we obtain
\begin{eqnarray}
p=-\frac{7}{1200}\frac{H_0^2c^2}{8\pi G}\left(\frac{3}{2}H_0t \right)^{-\frac{2}{3}}
\overline{(\triangle B)^2}\;.
\end{eqnarray}
This negative pressure is related to the gravitational attraction of
the underlying inhomogeneities rather than a cosmological constant. Expressing it in terms of the density
we get up to second order accuracy the equation of state
\begin{eqnarray}
p=-\frac{7c^2}{4800}\left(\frac{H_0^4}{3(\pi G)^2}\right)^{\frac{1}{3}}
\overline{(\triangle B)^2}\rho^{\frac{1}{3}}\;.
\end{eqnarray}

\begin{figure}[h]
\begin{flushleft}
\includegraphics{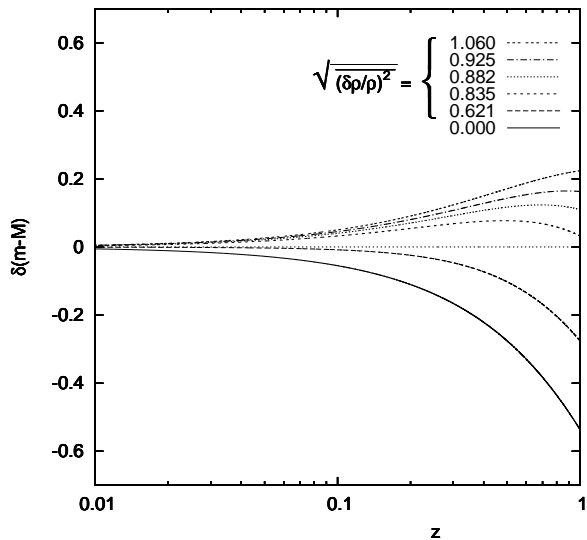}
\end{flushleft}
\caption{Differences in distance modulus versus the redshift. Different curves correspond to
 different strengths of relative density fluctuations $\sqrt{\overline{(\delta \rho /\rho)^2}}$.}
\label{fig2}
\end{figure}

In order to have a tentative comparison with supernova measurements, we calculate the luminosity distance
$d_L$ versus the redshift $z$ (Hubble diagram) by approximating the inhomogeneous universe with a homogeneous one, which
expands according to the corrected scale factor. 
Differences in distance modulus $m-M=5\log_{10}(d_L/10pc)$ (compared to an empty universe,
$d_L({\rm empty})=z+z^2/2$) are displayed in Fig.\ref{fig2}. This is to be compared with Figs.10-11. in \cite{Schmidt:1998}
or Figs.4-5. in \cite{Riess:1998cb}. Note that in the applied approximation curves belonging to higher values of 
relative density fluctuations converge to an upper bound which is roughly identical with the uppermost curve displayed.


Finally, we calculate the age of the universe according to Eq.(\ref{scale}). By solving the equation 
$\dot R(t_0)/R(t_0)=H_0$ we get 

\begin{eqnarray}
t_0=\frac{2}{3H_0}\frac{1+\frac{1}{2}\overline{\left(\frac{\delta\rho}{\rho}\right)^2_0 }}{1+\frac{1}{6}\overline{\left(\frac{\delta\rho}{\rho}\right)^2_0}}\;.
\end{eqnarray}
Assuming $\overline{\left({\delta\rho}/{\rho}\right)^2_0} \approx 1 $ it yields $t_0=\frac{9}{7}\frac{2}{3H_0}\approx 1.3 
\times \frac{2}{3H_0}$.

Some remarks are in order. The averaging procedure we used is not quite unambiguous. We might have included
e.g. the square root of the spatial metric. In this special case we obtain a further positive contribution
to the coefficient of the $t^2$ term in the scale factor. Nevertheless, it is advisable to consider 
directly the physical distance between two points as a more significant quantity. For points that are not very far
from each other, one has
$l=\sqrt{-g_{\alpha \beta}\Delta x_\alpha \Delta x_\beta}$,
or, inserting the previous expansions for $g_{\alpha \beta}$ ,
\begin{eqnarray}
l&\approx& l_0-\frac{1}{2 l_0}g_{\alpha \beta}^{(1)}\Delta x_\alpha \Delta x_\beta\nonumber\\
&-&\frac{1}{2 l_0}g_{\alpha \beta}^{(2)}\Delta x_\alpha \Delta x_\beta
-\frac{1}{8 l_0^3}\left(g_{\alpha \beta}^{(1)}\Delta x_\alpha \Delta x_\beta\right)^2
\end{eqnarray}
with
$l_0=\sqrt{-g_{\alpha \beta}^{(0)}\Delta x_\alpha \Delta x_\beta}=R^{(0)}\sqrt{\Delta x_\alpha^2}$.
By fixing the coordinate distance $d_0=\sqrt{\Delta x_\alpha^2}$ and averaging over both points  
one obtains for sub-horizon-sized perturbations
\begin{eqnarray}
l=d_0\left(R^{(0)}+\frac{3}{4000}\overline{\left(\triangle B\right)^2}\; H_0\; c\; t^2 \right)
\end{eqnarray}
where the second term indicates accelerating expansion. 

As for the Hubble diagram, a more precise calculation would take into account the inhomogeneities of the metric
along light trajectories, too, not only in the global scale factor. Such a procedure would need the complete second order 
perturbation term, including the spatial dependence. Note that a nonperturbative approach to this problem has been 
published in Ref.\cite{Linder}. As the spatial average of the second order term has proved to be large,
it may happen (if the variance is also large) that the actual Hubble diagram broadens to a strip. 
This would mean that the rather scattered measurement points follow a systematics and are not due simply to
experimental error.   
  
In conclusion, we have demonstrated that inhomogeneities essentially influence the time evolution of the late universe.
Our results may even need further corrections, since the terms treated perturbatively proved to be quite large. 
Nevertheless, it is remarkable that accelerating expansion may result from inhomogeneities, without assuming
a cosmological constant, quintessence, or modified Einstein equations.

\begin{acknowledgments}

The authors are indebted to Prof.~Zolt\'an Perj\'es, Prof.~Zal\'an Horv\'ath and Prof.~K\'aroly Nagy for their continued interest, several valuable remarks and stimulating discussions. We also thank Prof.~Amos Ori and Prof.~Hideo Kodama for an intriguing discussion.
This work has been supported by the OTKA grants T031724 and T043582. 
\end{acknowledgments}

\end{document}